\documentclass[prb,twocolumn,aps,showpacs]{revtex4}
\usepackage{graphics,graphicx,epsfig}
\begin{document}
\title{Scaling in reversible submonolayer deposition}
\author{T. J. Oliveira${}^{1,(a)}$ and F. D. A. Aar\~ao Reis${}^{2,(b)}$
\footnote{a) Email address: tiago@ufv.br\\
b) Email address: reis@if.uff.br}}
\affiliation{${}^{1}$ Departamento de F\'isica, Universidade Federal de Vi\c cosa, 36570-000,
Vi\c cosa, MG, Brazil\\
${}^{2}$ Instituto de F\'\i sica, Universidade Federal Fluminense, Avenida Litor\^anea s/n,
24210-340 Niter\'oi RJ, Brazil\\
}

\date{\today}

\begin{abstract}
The scaling of island and monomer density, capture zone distributions (CZDs), and island size
distributions (ISDs) in reversible submonolayer growth was studied using the Clarke-Vvedensky model.
An approach based on rate-equation results for irreversible aggregation
(IA) models is extended to predict several scaling regimes in square and triangular lattices, in agreement
with simulation results.
Consistently with previous works, a regime \textit{I} with fractal islands is observed at low temperatures,
corresponding to IA with critical island size $i=1$, and a crossover to a second regime appears as
the temperature is increased to $\epsilon R^{2/3}\sim 1$, where $\epsilon$ is the single bond detachment
probability and $R$ is the diffusion-to-deposition ratio.
In the square (triangular) lattice, a regime with scaling similar to IA with $i=3$ ($i=2$) is observed
after that crossover.
In the triangular lattice, a subsequent crossover to an IA regime with $i=3$ is observed, which is explained
by the recurrence properties of random walks in two dimensional lattices, which is beyond the
mean-field approaches.
At high temperatures, a crossover to a fully reversible regime is observed,
characterized by a large density of small islands, a small density of very large islands, and
total island and monomer densities increasing with temperature, in contrast to IA models.
CZDs and ISDs with Gaussian right tails appear in all regimes for $R\sim {10}^7$ or larger,
including the fully reversible regime, where the CZDs are bimodal. This shows that the
Pimpinelli-Einstein (PE) approach for IA explains the main mechanisms for the large islands to
compete for free adatom aggregation in the reversible model, and may be the reason for its
successful application to a variety of materials and growth conditions.
\end{abstract}

\maketitle

\section{Introduction}
\label{intro}

The morphology and physical properties of thin films and multilayers are strongly related to
the initial stages of their formation \cite{ohring,pimpinelli,venablesbook}. It motivated a
large number of works on modeling the submonolayer growth regime \cite{etb,venables1973,ratsch},
in which a single incomplete layer of adatoms is being formed.
These models usually consider a set of fundamental processes (diffusion, aggregation, etc)
in order to find activation energies and related quantities for a particular deposition
process or with the aim of investigating universal features valid for several materials and techniques.

The simplest models consider irreversible aggregation (IA) of atoms to islands
of size larger than a critical value $i$.
Rate-equation (RE) approaches predict the scaling of island and monomer densities ($N_{isl}$, $N_1$)
\cite{etb,venables1973} for any $i$ and show good agreement with simulation data,
usually performed for $i$ ranging between $1$ and $3$ \cite{etb}.
The shapes of the island size distributions (ISDs) were proposed by different approximation schemes
\cite{etb,venables1973,ratsch,af1995,amar2001,mb} and are still subject of debate.
A recent advance by Pimpinelli and Einstein (PE) \cite{pe} proposed that the capture zone
distributions (CZDs) are described by the Wigner surmise (WS) from random matrix theory \cite{rmt}
[the capture zone (CZ) of an island is defined as the area in which a diffusing adatom is more likely
to attach to that island than to any other one]. After an initial controversy \cite{shi,li,pereply},
that proposal was supported by simulation results
for various island shapes after suitable rescaling \cite{tiago2011}, under conditions of
high temperatures and low coverages. IA models are expected to model real systems in
temperatures sufficiently low to neglect atom detachment from islands, but this
condition is not obeyed in many real systems, even at room temperature. For this reason,
suitable choices of the size $i$ are frequently combined with internal restructuring mechanisms 
to reconcile experimental data and IA predictions \cite{af1995,ruiz,wu,shiqin,zheng}.

On the other hand, a smaller number of works have addressed scaling properties of models
with reversible aggregation (RA) \cite{ratschPRL,ratsch1995,bales,bartelt1995,mehl,furman,ramadan}.
Works on square lattices, which simulate metal $(100)$ homoepitaxy, have shown a
crossover between IA scaling regimes with $i=1$ and $i=3$ \cite{ratsch1995,bartelt1995}.
Partially reversible models allowing only single-bond detachment were studied using
simulations and improved RE approaches and confirmed the crossover scaling
between those regimes \cite{af1997,popescu}.
The work on triangular lattices, which is related to $(111)$ epitaxy, shows a crossover between regimes
with $i=1$ and $i=2$ \cite{af1997,popescu}.

In this paper, we study the scaling properties of island density, CZDs, and ISDs
in a model of fully reversible island growth, using numerical simulations in square and
triangular lattices and a scaling approach based on RE results for IA models.
We consider a canonical bond-counting model of Clarke-Vvedensky
\cite{cv} type, in which the activation energy for an adatom hop to a neighboring site
depends only on the number of nearest neighbors at its initial position, with the
temperature $T$ kept fixed during the growth.
It respects detailed balance conditions, in contrast to IA.
For a fixed coverage $\theta$ and increasing deposition temperature,
the islands evolve from a fractal shape (typical of IA with no island restructuring)
to a compact shape dominated by processes of adatom detachment and reattachment, and
the island density continuously decrease. The previously studied crossovers of IA regimes are reproduced,
but an additional crossover from $i=2$ to $i=3$ in triangular lattices is found.
At high temperatures, a nontrivial regime where the island density increases with temperature is
observed, with a large density of small clusters coexisting with large compact islands.
The Gaussian right tail of CZDs predicted by the PE approach is observed in all regimes for small
coverages and sufficiently high temperatures, even when the CZDs are bimodal.

The rest of this work is organized as follows. In Sec. \ref{model}, we present the RA model
and summarize previous theoretical results for IA and RA.
In Sec. \ref{square}, we present the simulation results and a scaling approach in square lattices.
In Sec. \ref{triangular}, that discussion is extended to triangular lattices.
Sec. \ref{conclusion} summarizes our results and presents our conclusions.

\section{Basic definitions and theoretical approaches}
\label{model}

In the RA model, 
the adsorbed atoms can occupy sites of a square or triangular lattice, with the maximum
of one adatom per site. There is a random flux of $F$ atoms per site per unit time.
Adsorption is allowed only if the site of incidence is empty, so that a single layer
is formed on the substrate lattice. This condition is reasonable for small coverages.
The hopping rate of an adatom at a given site $\vec{r}$ is given by
$D=D_0\exp{\left( -E_{act}/k_BT\right)}$  (proportional to its surface diffusion coefficient),
with the activation energy $E_{act}=E_s+nE_N$, where $n$ is the number of adatoms in
nearest neighbor (NN) sites of $\vec{r}$, $E_s$ is an energy barrier for surface diffusion
of free adatoms, and $E_N$ is the bond energy for each NN.

The parameters $F=0.1 s^{-1}$, $D_0={10}^{13} s^{-1}$, and $E_s=1.3 eV$ were kept fixed in
most simulations. The last value is characteristic of $Fe/Fe(100)$ epitaxy \cite{ratschPRL}.
Some data were obtained with $F=1 s^{-1}$ and $10 s^{-1}$ in order to confirm general scaling predictions. 
Broad ranges of temperature were studied, thus the conclusions can be extended
to other systems by considering the diffusion-to-deposition ratio of free adatoms $R\equiv D/F$
as the main model parameter. Due to the large frequency $D_0$, one typically has $R\gg 1$,
even at low temperatures.

Several values of the bond energy $E_N$ are analyzed, and scaling approaches are facilitated by
using the detachment probability (binding factor) $\epsilon\equiv \exp{\left( -E_N/k_BT\right)}$.
Since $E_N\geq 0.3 eV$ is considered throughout this work and the largest experimentally
relevant temperatures do not exceed $k_BT\sim 0.1 eV$, the condition $\epsilon \ll 1$ always applies.
It is also important to observe that both $R$ and $\epsilon$ increase with temperature.

Simulations were performed in very large lattices, typically of lateral size $L=1024$,
with at least $100$ realizations up to coverage $\theta=0.4$ for each parameter set.
Comparison with results in $L=512$ for the largest coverages and temperatures shows no
significant finite-size effect.
Standard algorithms were able to provide accurate estimates of densities and distributions
up to large values of $R$ (however, extensions to the multilayer regime or to models
with a larger number of atomistic processes would benefit from special algorithms developed
for this class of model \cite{shim}).

In the IA models, islands with a number of atoms larger than the critical size $i$ are
stable. Moreover, islands of size $i$ are assumed to have a binding energy $E_i<0$.
During the deposition process, after a transient regime, islands grow by capture of
diffusing adatoms, with negligible formation of new islands. In this steady state,
rate-equation theory \cite{venables1973,etb} predicts that the stable island density scales as
\begin{equation}
N_{isl}\sim \theta^{1/\left( i+2\right)} \exp{\left[ -\beta E_i/\left( i+2\right)\right]} R^{-\chi}
\label{Nisl}
\end{equation}
and the monomer (free adatom) density scales as
\begin{equation}
N_1\sim \theta^{-1/\left( i+2\right)} \exp{\left[ \beta E_i/\left( i+2\right)\right]} R^{\chi -1} ,
\label{N1}
\end{equation}
where
\begin{equation}
\chi = i/\left( i+2\right) .
\label{chi}
\end{equation}
These results assume that the mean capture number of stable islands is independent of the coverage.

Information on the growth dynamics can also be extracted from
the probability densities of island size $s$, $Q(s)$, and of CZ area $x$, $P(x)$ (ISD and CZD,
respectively). The ISD follows the scaling form
\begin{equation}
Q(s)=\frac{1}{\langle s\rangle}
f{\left( \frac{s}{\langle s\rangle}\right)} ,
\label{scalingtrad}
\end{equation}
where $f$ is a scaling function, and an equivalent scaling form applies to $P(x)$.
Alternatively, scaling with the variance
$\sigma_x\equiv {\overline{{\left( x-\langle x\rangle\right)}^2}}^{1/2}$ may be used as
\begin{equation}
P(x) = \frac{1}{\sigma_x} g{\left( \frac{x-\langle x\rangle}{\sigma_x}\right)} .
\label{scalingvariance}
\end{equation}
This procedure was previously used with IA models \cite{tiago2011} and is helpful in other problems,
such as scaling of roughness distributions of thin film growth models \cite{intrinsic}.

The best know approach to predict the shape of CZD in IA is that of PE, which proposes the
CZD to be described by the WS
\begin{equation}
P_\beta(z) = a_\beta z^{\beta} \exp{\left( -b_\beta z^2\right)} ,
\label{ws}
\end{equation}

\begin{figure*}[!t]
\includegraphics[width=16.7cm]{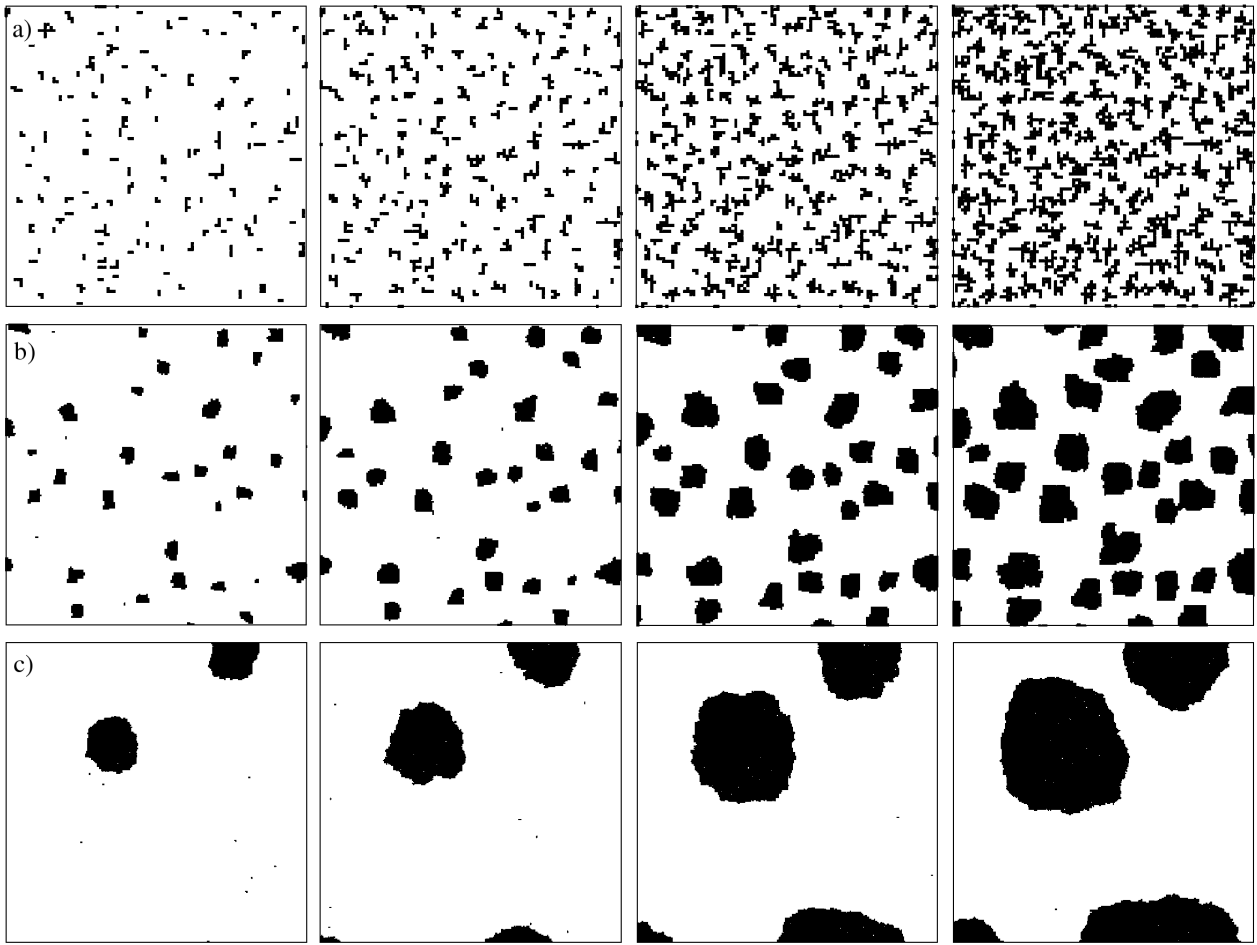}
\caption{Square lattice islands for (a) $T=650 K$, (b) $T=900 K$, and (c) $T=1200 K$.
The panels are $100\times100$, $200\times200$, and $300\times300$ cuts of systems of size $L=1024$, respectively.
The coverages are $\theta=0.05$, $0.10$, $0.20$, and $0.30$ from left to right.}
\label{fig1}
\end{figure*}

where $z\equiv x/\left\langle x\right\rangle $, $\beta = \frac{2}{d}\left( i+1\right)$,
$d$ is the substrate dimension ($d=2$ in
the present work) and the parameters $a_\beta$ and $b_\beta$ are determined by normalization conditions.
This proposal follows from the phenomenological argument that the CZD
can be extracted from a Langevin equation representing the competition of neighboring islands
for adatom aggregation. Several experimental works have already shown agreement of CZDs with the WS,
such as growth of para-sexiphenyl islands \cite{lorbek,potocar},
$Cu$ deposition with impurities \cite{sathiyanarayanan}, pentacene island growth with
impurities \cite{conrad}, $InAs$ quantum dot
growth on $GaAs$ \cite{arciprete2010}, and $C_{60}$ deposition on $SiO_2$ films \cite{groce}.

In Ref. \protect\cite{tiago2011}, a scaling approach was used to predict
the decay of the right tail of the ISD from the Gaussian tail of the CZD (Eq. \ref{ws}) and
the island shape (point, fractal or square).

\section{Square lattice}
\label{square}

\subsection{Scaling of island and monomer density}
\label{squaredensity}

Figs. 1a-c illustrate the island evolution for $E_N=0.4 eV$ and temperatures representative of
different scaling regimes. Fig. 2a shows the temperature evolution (parametrized by $R$) of
the island density and monomer density for coverage $\theta =0.1$ and several values of $E_N$.

For very low temperatures ($R\lesssim {10}^3$), both densities are nearly temperature-independent,
and the system behaves as in random sequential adsorption (RSA) without diffusion \cite{privman},
as discussed in Ref. \protect\cite{tiago2012}. 

The subsequent low temperature regime, hereafter called regime \textit{I}, is characterized by the
decrease of $N_{isl}$ and $N_1$ with $R$, as shown in Fig. 2a. In this regime, the probability of
detachment of an aggregated adatom from an island is negligible during the time interval necessary
for the aggregation of a new adatom to its neighborhood. This regime is equivalent to IA with $i=1$.
Using Eq. (\ref{Nisl}) for fixed coverage, one has 
\begin{equation}
N_{isl}\sim R^{-1/3} , \quad N_1\sim R^{-2/3} \quad \quad (I)
\label{NislI}
\end{equation}

\begin{figure}[!t]
\includegraphics[width=8cm]{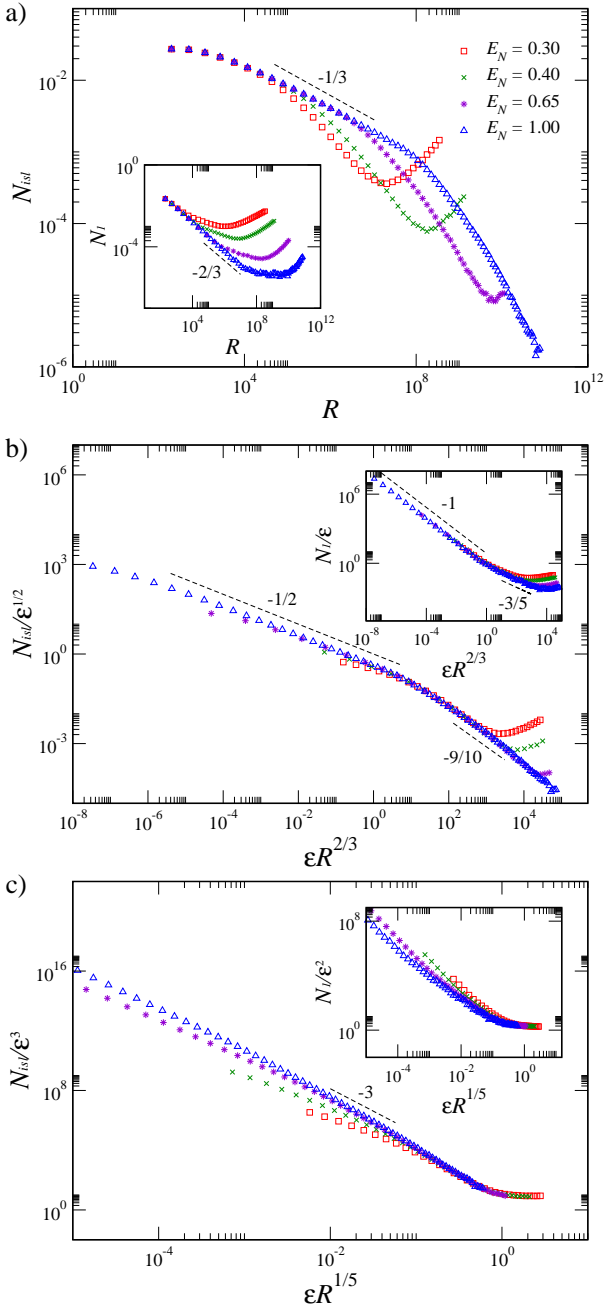}
\caption{(Color online) Scaled island densities
for $\theta=0.10$ and several binding energies in the square lattice.
Dashed lines indicate local slopes. The insets show scaled monomer densities.}
\label{fig2}
\end{figure}

This scaling is highlighted in Fig. 2a. The longer regimes with $i=1$ scaling are observed for
$E_N=1.00 eV$; for $5\times{10}^4\leq R\leq {10}^7$ (nearly three decades), fits of the
data give $N_{isl}\sim R^{-0.33(1)}$ and $N_1\sim R^{-0.67(3)}$, which are consistent with Eq. (\ref{NislI}).

The average detachment time of a singly bonded atom is
\begin{equation}
\tau_1\sim 1/\left( D\epsilon\right) .
\label{tau1}
\end{equation}
In a mean-field approach (RE), the time for a free adatom to encounter that aggregated atom is
\begin{equation}
\tau_{ag}\sim 1/\left( D N_1\right) .
\label{tauag}
\end{equation}
Regime  \textit{I} is characterized by $\tau_{ag}\ll\tau_1$, so that adatom bonds are effectively
stable.

A crossover to a second regime is observed when $\tau_1$ matches $\tau_{ag}$
[Eqs. (\ref{tau1}) and (\ref{tauag})], as explained in Refs. \protect\cite{bartelt1995,ratsch1995,af1997}
(see also Sec. 8.5 of Ref. \protect\cite{etb}). A crossover variable is defined as 
\begin{equation}
Y_1\equiv\epsilon R^{2/3} ,
\label{Y1}
\end{equation}
with $Y_1\ll 1$ in regime \textit{I}. Regime \textit{II} begins with $Y_1\sim 10$.
From Eqs. (\ref{NislI}) and (\ref{Y1}), the crossover scaling for island density
and monomer density is
\begin{equation}
N_{isl}\sim \epsilon^{1/2}F_1\left( Y_1\right) , \quad N_1\sim \epsilon G_1\left( Y_1\right) , 
\label{NislcrossI}
\end{equation}
where $F_1$ and $G_1$ are scaling functions.
Regime \textit{II} is effectively characterized by $i=3$, since detachment of doubly bonded
atoms of square islands (size $s=4$) occurs in much longer timescales.
Thus, for $Y_1>10$ but not very large,
Eqs. (\ref{Nisl}), (\ref{N1}), and (\ref{NislcrossI}) give
\begin{equation}
N_{isl}\sim \epsilon^{-2/5}R^{-3/5} , \quad N_1\sim \epsilon^{2/5} R^{-2/5} \quad\quad (II_{initial}) .
\label{NislIIinitial}
\end{equation}
This result is consistent with a binding energy $E_i=2E_N$ for the critical island with $i=3$ adatoms.

Fig. 2b illustrates the crossover scaling with the collapse of data in regimes \textit{I} and \textit{II}
for various $E_N$.
Again, the longest regime with $i=3$ scaling is observed for $E_N=1.00 eV$; fits of the
data for ${10}^2\leq \epsilon R^{2/3}\leq {10}^4$ (nearly two decades in $R$) give
$N_{isl}\sim R^{-0.92(4)}$ and $N_1\sim R^{-0.56(5)}$, which are consistent with 
Eqs. (\ref{NislI}) and (\ref{NislIIinitial}) (see the predicted slopes in Fig. 2b).

For $E_N=1.0$, Fig. 2a shows that the slope of the $N_{isl}\times R$ plot slowly increases in
regime \textit{II}. This occurs because the binding energy $E_i$ becomes negligible at the
end of this regime, so that $N_{isl}$ and $N_1$ scale tend to scale only with $R$, as
\begin{equation}
N_{isl}\sim R^{-3/5} , \quad N_1\sim  R^{-2/5} \quad\quad (II_{final}) 
\label{NislIIfinal}
\end{equation}
[the same power laws of Eq. (\ref{NislIIinitial}) excluding the $\epsilon$-dependence].
The same evolution of slopes is observed in all effective IA regimes observed in the RA model
(see Sec. \ref{triangular}).

As the temperature increases, a crossover to a new scaling regime is expected by two reasons:
rapid detachment of doubly bonded
atoms from islands with $s\geq 4$, which occurs in a characteristic time
\begin{equation}
\tau_2\sim 1/\left( D\epsilon^2\right) ,
\label{tau2}
\end{equation}
or slow atom deposition in the CZ, which occurs in a characteristic time
\begin{equation}
\tau_{dep}\sim N_{isl}/F .
\label{taudep}
\end{equation}

\begin{figure}[!t]
\includegraphics[width=8cm]{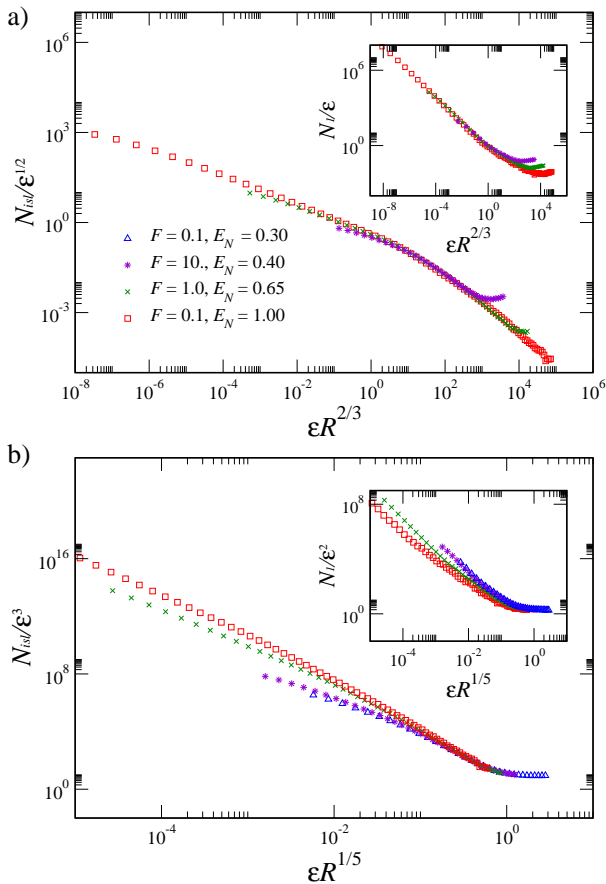}
\caption{(Color online) Scaled island densities
for $\theta=0.10$ and different deposition rates and binding energies in the square lattice.
The insets show scaled monomer densities.}
\label{fig3}
\end{figure}

In the steady state of IA, $\tau_{dep}$ [Eq. (\ref{taudep})] is of the same order of
$\tau_{ag}$ [Eq. (\ref{tauag})]. Thus,
regime \textit{II} is expected to end as $\tau_2$ [Eq. (\ref{tau2})] matches $\tau_{ag}$, i. e., when
$\epsilon R^{1/5}\sim 1$. This defines a second crossover variable as
\begin{equation}
Y_2\equiv\epsilon R^{1/5} .
\label{Y2}
\end{equation}
Crossover scaling follows from Eq. (\ref{NislIIfinal}) as
\begin{equation}
N_{isl}\sim \epsilon^3 F_2\left( Y_2\right) , \quad N_1\sim \epsilon^2 G_2\left( Y_2\right) . 
\label{NislcrossII}
\end{equation}

Fig. 2c illustrates the crossover scaling for various $E_N$, with an excellent
collapse of $N_{isl}$ data. Some deviations appear for $N_1$, probably
due to scaling corrections. The expected slope $-3$ before the crossover, predicted by Eq. (\ref{NislIIfinal}),
is also shown in Fig. 2c. The fits of the data for $E_N=1.0 eV$ and $E_N=0.65 eV$, in the range
${10}^{-2}\leq \epsilon R^{1/5}\leq {10}^{-1}$ (nearly one decade in $R$), give slopes
$-3.3(2)$ and $-3.2(1)$, respectively, both close to the theoretical value. This gives additional support
to the proposed evolution in regime \textit{II}, to an $\epsilon$-dependent to an $\epsilon$-independent
scaling.

After the crossover, a fully reversible regime \textit{III} is attained, completely different from IA.
Fig. 2a shows that the island density and the monomer density increase with temperature in this regime,
in striking contrast with regimes \textit{I} and \textit{II} and all IA models.
Fig. 1c shows that it is characterized by a very large density
of small islands, mainly isolated adatoms, and a small density of very large islands.
As time evolves, these large islands grow at the expense of the small ones, i. e., there is
island ripening \cite{comsa,grabow}.

In regime \textit{III},
the density of small islands is much larger than the density of large islands. This is an expected evolution
from regime \textit{II} because $N_1$ decreases with temperature slower than $N_{isl}$
[Eq. (\ref{NislIIinitial})].
This trend is expected for any regime with $i>2$, which gives $\chi>1/2$, and will also be shown
in the triangular lattice.

The high-temperature scaling in regime \textit{III} is only $\epsilon$-dependent, with
\begin{equation}
N_{isl}\sim \epsilon^{3} , \quad N_1\sim \epsilon^2 \quad \quad (III) .
\label{NislIII}
\end{equation}
Indeed, since movement of free adatoms is very fast, only the attachment-detachment dynamics is
important to determine the densities.

This steady state regime can be predicted by noting that double bonding is the
minimal possible bonding for an island in the square lattice, i. e. at least four atoms of a large
island have two bonds. Equating aggregation and detachment rates obtained from
Eqs. (\ref{tauag}) and (\ref{tau2}), we get the above relation for $N_1$. Since regime \textit{III}
takes place as the evolution of an effective $i=3$ IA regime (also related to lattice structure),
where $N_{isl}\sim N_1^{3/2}$, we obtain the above relation for $N_{isl}$.

In order to search for possible effects of deposition rate ($F$) or amplitude of hopping rates
($D_0$), Fig. 3 compares the crossover scaling  (a) from regimes \textit{I} to \textit{II}
and (b) from \textit{II} to \textit{III}, obtained for $0.1 s^{-1} \leq F \leq 10 s^{-1}$
and different binding energies $E_N$. The good data collapses in Fig. 3 show the general validity
of the above scaling relations.

\subsection{Capture-zone and island size distributions}
\label{squaredistr}

Despite the effective IA scaling of regime \textit{I}, the CZDs usually do not match the WS with $i=1$
($P_{2}$) for the values of $E_N$ and temperatures considered here.
Fig. 2a shows that regime \textit{I} typically occurs with $R\lesssim {10}^7$.
This range is intermediate between low adatom mobility (RSA behavior) and high adatom
mobility (PE behavior), thus CZDs show a crossover scaling, as shown in Ref. \protect\cite{tiago2012}.
Similar crossover is observed in the ISDs.
For instance, considering $E_N=1.0$, for $R \lesssim 10^5$ we observe simple exponential tails of CZDs
(RSA) and for $R\sim {10}^7$ they become Gaussian (PE), but still not collapsing with the WS.

\begin{figure}[!h]
\centering
\includegraphics[width=8cm]{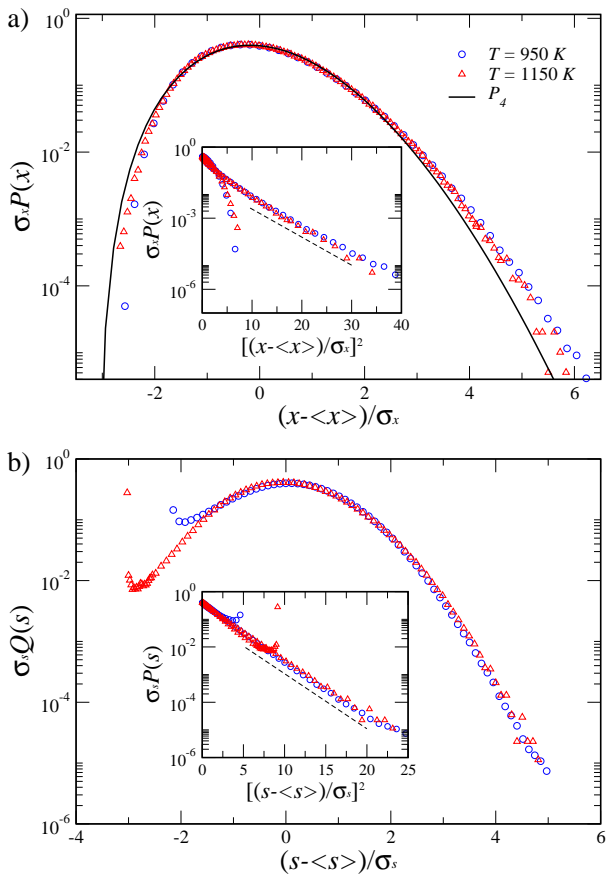}
\caption{(Color online) Scaled (a) CZDs and (b) ISDs for the energy $E_{N} = 0.65$ and temperatures
$T=950K$ ($R\sim 10^{7}$) and $T=1150K$ ($R\sim 2\times 10^{8}$). The insets show the same scaled
distributions with squared abscissas. The dashed lines highlight the Gaussian decay of the right tails.}
\label{fig4}
\end{figure}

\begin{figure}[!h]
\centering
\includegraphics[width=7cm]{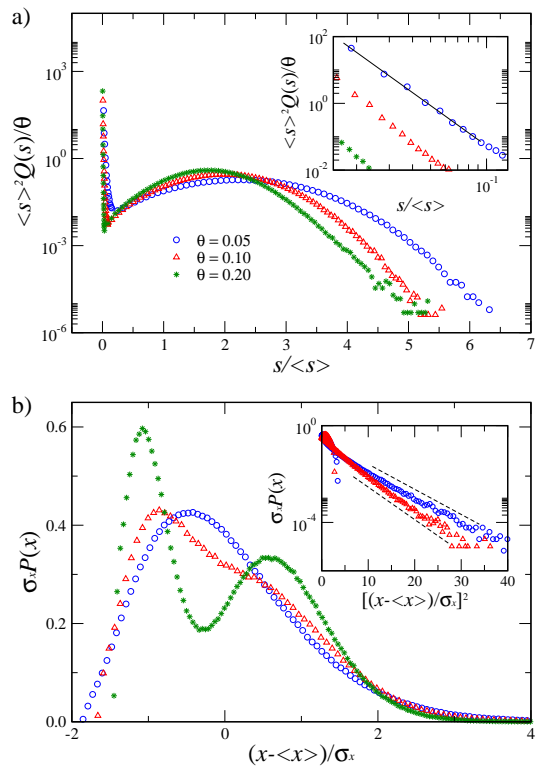}
\caption{(Color online) (a) Scaled ISDs for $E_{N}=0.30$, $T=950K$ ($R\approx 10^7$) and several coverages.
The inset shows the same data in a log-log scale and the line has slope $-4$.
(b) Scaled CZDs for the same parameters of (a). The inset shows the same distributions with squared abscissas.
The dashed lines highlight the Gaussian decays of the right tails.}
\label{fig5}
\end{figure}

The CZDs in regime \textit{II} are not well fitted by the WS with $i=3$ ($P_4$ in Eq. \ref{ws}),
probably because this is not a true IA process, as shown in Fig. 4a.
However, as the temperature increases, the CZD gets closer to that WS.
This occurs considering scaling with the average [Eq. (\ref{scalingtrad})]
or with the variance [Eq. (\ref{scalingvariance})].
On the other hand, Fig. 4a shows that the tails of the CZDs are Gaussian, similarly to IA models
\cite{tiago2011}.
This trend fails only for very small $E_N$ because regime \textit{II} is attained with small $R$
and crossover from RSA is present.

The ISD has features similar to those of IA models of compact islands (e. g. square islands),
as shown in Fig. 4b:
the left tail is high, indicating the presence of a large number of small islands, and
the right tail shows a Gaussian decay.
As the coverage increases, a crossover to simple exponential decay is observed in the CZDs and ISDs, 
similarly to the IA models \cite{tiago2012}.

Fig. 5a shows ISDs in regime \textit{III}, with left tails much higher than the peaks, confirming the
presence of a large density of free adatoms and small islands.
The left tail shows even-odd oscillations characteristic of loose-packed lattices (inset of Fig. 5a).
If the small islands were in near-equilibrium conditions, the Walton relation
\cite{venables1973,walton} $Q(s)\approx {N_1}^s$ would apply and an exponential decay of the left tail
would be observed. However, this is not the case: the inset of Fig. 5a shows an approximately
power-law decay, with large exponents that depend on the coverage, which suggests that more complex
mechanisms govern the small island dynamics.

In regime \textit{III}, $\tau_{dep}\ll\tau_2$, thus a small island will probably disappear before
a new atom is deposited in its CZ. The concept of CZ as the probable region for deposition of a new atom
to be captured by that island becomes irrelevant. Anyway, we also measured the CZDs in that regime and
observed that they are bimodal, as shown in Fig. 5b. The first peak corresponds
to small islands and the second peak corresponds to the large islands.
The bimodal CZDs are completely different from those of IA models, where distortions
of the monomodal shapes of ISDs and CZDs appear only for large coverages \cite{tiago2012}.
As $\theta$ increases, the first peak of the CZD moves to the left (smaller CZs for small islands)
and the second one moves to the right (larger CZs for large islands).

A surprising result in the inset of Fig. 5b is that the Gaussian right tail of the CZDs
is preserved for low coverages in regime \textit{III}. Thus, the competition of large islands for the capture of
diffusing adatoms is still that predicted by the PE approach in the fully reversible regime.
The small islands, which rapidly dissociate into isolated monomers, contribute
to maintain the competitive dynamics of the neighboring large islands.
The right tails of the corresponding ISDs are also Gaussian, following the trend of compact
islands already explained in Ref. \protect\cite{tiago2011}.

In experiments, very small islands ($s=2,3,\cdots$) may not be detected in
microscopy images, except if high resolution techniques are used.
If that is the case, the first peak of the bimodal CZDs may be lost,
but the Gaussian right tail will be preserved. This may suggest IA behavior in a system
with fully reversible dynamics.

\section{Triangular lattice}
\label{triangular}

Fig. 6a shows the $R$-dependence of the island density for various values of $E_N$.
It has the main features of the RA model in the square lattice, including the high temperature
crossover to the fully reversible regime, where $N_{isl}$ and $N_1$ increase with $R$.
However, some important features of RA are particular of the triangular lattice,
as discussed below. 

The crossover from a fractal island regime \textit{I} ($i=1$) to the first regime of compact islands is
equivalent to that in the square lattice, occurring when $\tau_1$ matches $\tau_{ag}$ and
leading to the crossover relations in Eq. (\ref{NislcrossI}). This is illustrated
in Fig. 6b. The subsequent regime is hereafter called \textit{IIa} and corresponds to IA with $i=2$,
i. e. with the smaller stable island being a triangle with $s=3$ adatoms, shown in Fig. 7a.
From Eq. (\ref{NislcrossI}) and the RE equations (\ref{Nisl}) and (\ref{N1}) for $i=2$,
regime \textit{IIa} has initial scaling as
\begin{equation}
N_{isl}\sim \epsilon^{-1/4}R^{-1/2} , \quad N_1\sim \epsilon^{1/4} R^{-1/2} \quad \quad (IIa_{initial}) .
\label{NislIIainitial}
\end{equation}
This trend is confirmed by the slope of the curves in Fig. 6b after the crossover.
Eq. (\ref{NislIIainitial}) is consistent with a binding energy of the critical island
$E_i=E_N$, which is reasonable for a two-adatom island.

As the temperature increases, regime \textit{IIa} evolves so that the critical island binding energy
becomes negligible and island and monomer densities scale as
\begin{equation}
N_{isl}\sim R^{-1/2} , \quad N_1\sim R^{-1/2} \quad \quad (IIa_{final}) .
\label{NislIIafinal}
\end{equation}
This evolution parallels that observed in the square lattice.

A crossover to another regime also occurs when the typical aggregation time to an island is
of the order of the detachment time $\tau_2$ of doubly-bonded atoms (Fig. 7a). Matching
Eqs. (\ref{tauag}) and (\ref{tau2}) using (\ref{NislIIafinal}), we obtain a crossover variable
\begin{equation}
Y_3\equiv\epsilon R^{1/4} 
\label{Y3}
\end{equation}
and crossover scaling as
\begin{equation}
N_{isl}\sim \epsilon^{2}F_3\left( Y_3\right) , \quad N_1\sim \epsilon^2 G_3\left( Y_3\right) . 
\label{NislcrossIII}
\end{equation}

\begin{figure}[!h]
\includegraphics[width=7cm]{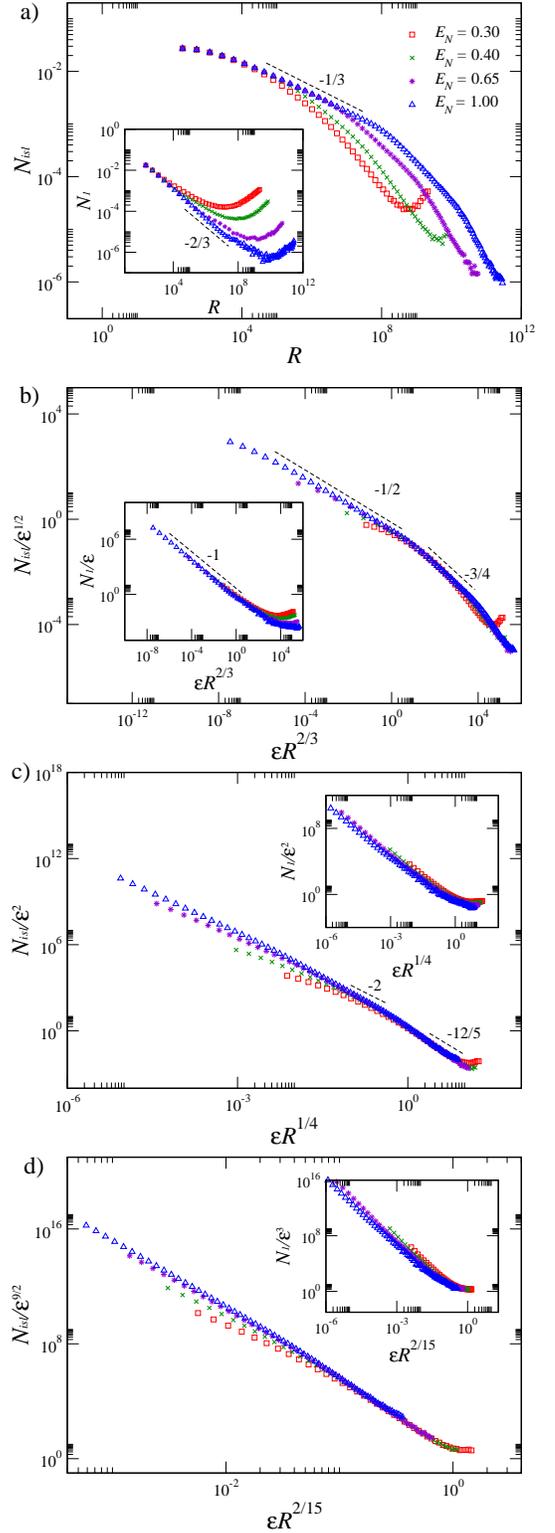}
\caption{(Color online) Scaled island densities for $\theta=0.10$ and several binding
energies in the triangular lattice.
Dashed lines indicate local slopes. The insets show scaled monomer densities.}
\label{fig6}
\end{figure}

In the square lattice, a similar crossover leads to regime 
\begin{figure*}[!t]
\centering
\includegraphics[width=8.5cm]{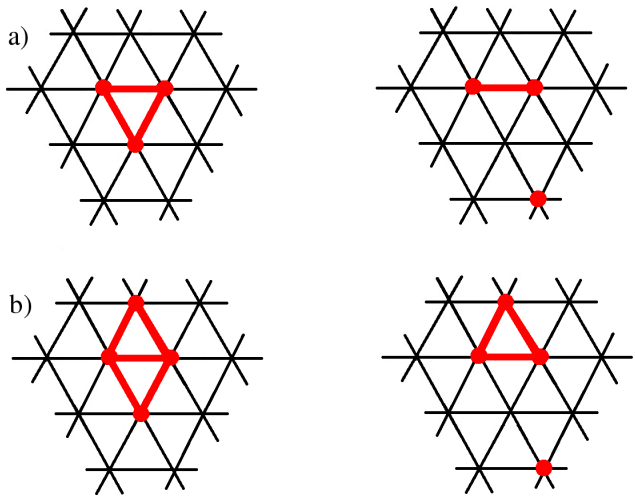}
\caption{(Color online) Illustration of stable islands of size (a) $s=3$ ($i=2$)
and (b) $s=4$ ($i=3$) on a triangular lattice before (left) and after (right) loosing a particle. }
\label{fig7}
\end{figure*}
\textit{III}.
However, here it leads to another regime with IA scaling due to the triangular lattice geometry.
The reason is that an island with $s=4$ atoms (Fig. 7b), is stable for a time much longer than
the island with $s=3$ atoms (Fig. 7a), as will be explained below.
Thus a regime with $i=3$ appears when $R$ (or $T$) increases, which is
hereafter called regime \textit{IIb}. Its scaling properties follow by matching
the $R$-dependence of Eqs. (\ref{Nisl}) and (\ref{N1}) for $i=3$ and that in Eq. (\ref{NislcrossIII}):
\begin{equation}
N_{isl}\sim \epsilon^{-2/5}R^{-3/5} , \quad N_1\sim \epsilon^{2/5} R^{-3/5} \quad\quad (IIb_{initial}) .
\label{NislIIbinitial}
\end{equation}

Fig. 6c illustrates this crossover scaling. For $E_N=1.00 eV$ and $E_N=0.65 eV$, regime
\textit{IIa} (final) is represented approximately by $5\times{10}^{-2}\leq \epsilon R^{1/4}\leq 5\times{10}^{-1}$
and regime \textit{IIb} (initial) is represented approximately by 
$2 \leq \epsilon R^{1/4}\leq {10}$, both corresponding to nearly one order of
magnitude in $R$. Linear fits of those data provide slopes $-2.2(1)$ and $-2.4(1)$, respectively. Both estimates are
in good agreement with the predicted slopes shown in Fig. 5c [from Eqs. (\ref{NislIIafinal}) and
(\ref{NislIIbinitial})].

Figs. 7a and 7b show the islands with $s=3$ and $s=4$ atoms loosing one adatom, which provide the
(critical) islands with $i=2$ and $i=3$, respectively. Both processes of loosing one adatom occur in a
characteristic time $\tau_2$ [Eq. \ref{tau2}].
The island with $i=2$ needs a time $\tau_1$ to break into two monomers,
while the island with $i=3$ needs a time $\tau_2$ to break into isolated monomers.
The addition of these times do not change the order of magnitude of the mean-field rate $1/\tau_2$.
However, the stability time of the remaining islands in Figs. 7a and 7b is affected
if the recurrence property of two-dimensional random walks is considered.

The probability that a random walker returns to the origin after $n$ steps is of order
$1/n^2$ \cite{montroll}. Thus, the probability that the island with $i=2$ breaks before the
return of the detached atom (Fig. 7a) is of order $\int_{D\tau_1}^{\infty}{1/n^2 dn}\sim \epsilon$.
Analogously, the probability that the island with $i=3$ breaks before the
return of the detached atom (Fig. 7b) is of order $\int_{D\tau_2}^{\infty}{1/n^2 dn}\sim \epsilon^2$.
Thus, the lifetime of the island with $i=3$ is $\epsilon^{-1}$ times larger
than the lifetime of the island with $i=2$. For $\epsilon\ll 1$, a factor $\epsilon^{-1}$ has to
be introduced in the characteristic time for an island with $s=4$ adatoms to break.

Following these ideas, we incorporate that factor in the mean-field approach
by stating that the effective lifetime for the critical island with $i=3$ is
\begin{equation}
\tau_3\sim \epsilon^{-1}\tau_2\sim 1/\left( D\epsilon^3\right) .
\label{tau3}
\end{equation}
This does not represent a failure of the RE theory, but a suitable form to extend its application.
We recall that the RE approach overestimates rates of several processes, as shown in simulation of IA models
\cite{etb,zhang,popescu}, but they are overestimated (and characteristic times
underestimated) by similar factors, so that the final scaling on $R$, $\theta$ and $E_i$ is correct.

The scaling of $N_{isl}$ and $N_1$ also evolve to $\epsilon$-independent forms in regime \textit{IIb} as the
temperature increases:
\begin{equation}
N_{isl}\sim R^{-3/5} , \quad N_1\sim R^{-3/5} \quad\quad (IIb_{final}) .
\label{NislIIbfinal}
\end{equation}

In regime \textit{IIb}, the monomer density decreases with temperature slower than island density. Thus,
it is expected to be followed by the fully reversible regime \textit{III}, similarly to the square lattice.
Matching $\tau_3$ with the aggregation time [Eq. (\ref{tauag})] for the $i=3$ regime, the crossover
scaling variable is defined as
\begin{equation}
Y_4\equiv\epsilon R^{2/15} .
\label{Y4}
\end{equation}
The crossover scaling is obtained considering Eq. (\ref{NislIIbfinal}):
\begin{equation}
N_{isl}\sim \epsilon^{9/2}F_4\left( Y_4\right) , \quad N_1\sim \epsilon^3 G_4\left( Y_4\right) . 
\label{NislcrossIV}
\end{equation}
Fig. 6d illustrates this crossover scaling (again with some deviations in the collapse of $N_1$
due to corrections to scaling).

After this crossover, the high-temperature, fully reversible regime \textit{III} is observed, with
the same basic features of the square lattice. The island and monomer densities scale only with
$\epsilon$ as
\begin{equation}
N_{isl}\sim \epsilon^{9/2} , \quad N_1\sim \epsilon^3 \quad\quad (III) ,
\label{NislIIItr}
\end{equation}
thus increasing with temperature.
This is also a steady state, nearly deposition independent, where $N_1$ follows from matching
of aggregation and detachment times [Eqs. (\ref{tauag}) and (\ref{tau3})] and $N_{isl} \sim N_1^{3/2}$,
a relation characteristic of the largest available critical island.

The features of ISDs and CZDs in the triangular lattice are similar to those of the square lattices,
including the Gaussian tails observed at high temperatures and extending to the fully reversible regime.

\section{Conclusion}
\label{conclusion}

We analyzed the scaling of island density, CZDs, and ISDs in a model of fully reversible island
growth where the activation energy for an adatom hop depends only on the number of NNs at its initial
position (Clarke-Vvedensky model). A scaling approach based on rate-equation results for IA models
was presented and supported by numerical simulations in square and triangular lattices.

At low temperatures, a regime \textit{I} with fractal islands is observed, corresponding to IA with critical
island size $i=1$. CZDs and ISDs are usually intermediate between those of RSA (negligible diffusion
of free adatoms) and those of PE theory (competition of large islands for rapidly moving adatoms),
except for large values of the binding energy $E_N$, where the regime is extended up to $R\sim{10}^7$
and a nearly Gaussian right tail of CZDs appears.

A crossover to a second regime is always observed as the reduced model parameters satisfy
$\epsilon R^{2/3}\sim 1$, in agreement with previous works. In the square (triangular) lattice,
a regime with scaling similar to IA with $i=3$ ($i=2$) is observed after the crossover. The initial
dependence of island density and monomer density with $\epsilon$ disappears as the temperature increases,
so that those quantities scale only with $R$ close to the subsequent crossover.

In the triangular lattice, there is a crossover to a third IA regime, with $i=3$. This is explained
by the recurrence properties of random walks in two dimensions, since the detachment times of doubly
bonded atoms in islands with $3$ and $4$ atoms are the same and that feature could not be predicted
solely with a mean-field approach.

Although the CZDs shows some deviations from the WS in these regimes with $i>1$, their right
tails are typically Gaussian, indicating that the PE theory is reasonable to represent the competition
of large islands for free adatom capture. Exceptions appear for small values of $E_N$, where 
those regimes may appear for small $R$.

A final crossover to a fully reversible regime \textit{III} is observed as $Y_2 \sim 1$ ($Y_4 \sim 1$) in
the square (triangular) lattice. In contrast to IA scaling, this regime is characterized by an
increase of the island density and of the monomer density with temperature, for fixed coverage.
A large density of small islands coexists with a small density of very large islands.
The corresponding ISDs show a high left tail and the CZDs are bimodal.
The Gaussian right tails of CZDs and ISDs are also observed, despite the reversible
nature of the aggregation, and are explained because the small islands do not change the
main mechanisms for the large islands to compete for free adatom aggregation.

These results show that the Gaussian right tails of CZDs are kept in reversible aggregation, as
predicted by the PE approach for IA. The same occurs with ISDs.
In experiments where the resolution of imaging techniques is not high, detection of small islands
may be difficult and even the high temperature regimes may have ISDs and CZDs similar to those of IA.
This may explain the success of PE predictions to describe a variety of experimental results,
for various materials and growth conditions
\cite{lorbek,potocar,sathiyanarayanan,conrad,arciprete2010,groce}.

\acknowledgments

The authors acknowledge support from CNPq, FAPEMIG and FAPERJ (Brazilian agencies).


\end{document}